\documentclass[12pt,preprint]{aastex}

\newcommand{\be}{\begin{equation}}
\newcommand{\ee}{\end{equation}}

\begin{document}

\shorttitle{Diffusion of cosmic-rays and the GLAST}

\shortauthors{\textsc{Rodriguez Marrero, Torres, de Cea del Pozo, Reimer, Cillis}}

\title{Diffusion of cosmic-rays and the Gamma-ray Large Area Telescope: \\ Phenomenology at
the 1--100 GeV regime}

\author{Ana Y. Rodriguez Marrero\altaffilmark{1}, Diego F. Torres\altaffilmark{1,2},  Elsa de Cea del Pozo\altaffilmark{1}, Olaf Reimer\altaffilmark{3}, \& Anal\'{\i}a N. Cillis\altaffilmark{4}}

\altaffiltext{1}{Institut de Ciencies de l'Espai (IEEC-CSIC)
  Campus UAB, Fac. de Ciencies, Torre C5, parell, 2a planta, 08193
  Barcelona,  Spain. E-mails: arodrig@ieec.uab.es, decea@ieec.uab.es}
\altaffiltext{2}  {Instituci\'o Catalana de Recerca i Estudis Avan\c{c}ats (ICREA), Spain. E-mail: dtorres@ieec.uab.es }
\altaffiltext{3}{Stanford University, W.~W. Hansen Experimental
  Physics Lab (HEPL) and KIPAC, Stanford, CA 
  94305-4085, USA. E-mail: olr@stanford.edu}
\altaffiltext{4}{NASA Goddard Space Flight Center,
Greenbelt, MD 20771. E-mail: analia.n.cillis@nasa.gov }

\begin{abstract}
This paper analyzes astrophysical scenarios that may be detected at the upper end of the energy range of  the  Gamma Ray Large Area Space Telescope (GLAST), as a result of cosmic-ray (CR) diffusion in the interstellar medium (ISM). Hadronic processes are considered as the source of $\gamma$-ray photons from localized molecular enhancements nearby accelerators. Two particular cases are presented: a) the possibility of detecting spectral energy distributions (SEDs) with maxima above 1 GeV, which may be constrained by detection or non-detection at very-high
energies (VHE) with observations by ground-based Cerenkov telescopes, and b) the possibility of detecting V-shaped, inverted spectra, due to confusion of a nearby (to the line of sight) arrangement of
accelerator/target scenarios  with different characteristic properties. We show
that the finding of these signatures (in particular, a peak at the 1--100 GeV energy region) is indicative for an identification of the
underlying mechanism producing the $\gamma$-rays that is realized by nature: which accelerator (age and relative position to the target cloud) and under which diffusion properties CR propagate.
\end{abstract}



\section{Introduction}

1506. This is the number of cosmic photons with energies above 10 GeV that were detected during the 9 years lifetime of the Energetic Gamma Ray Experiment Telescope (EGRET, Thompson et al. 2005).  
{  Out of this number of photons, 187 photons were found within 1 degree of sources that are listed in the Third EGRET Catalog (Hartman et al. 1999) and can be plausible related to the more energetic extent of the cataloged EGRET sources. The majority of the remaining photons correspond to diffuse Galactic and extragalactic radiation, albeit this conclusion is based on the similarity of their spatial and energy distributions with the diffuse model. }
No significant time clustering nor source shining only at such high energies was detected. The scarcity of these detections represents the last electromagnetic window that remains to be opened between already explored energy ranges.
EGRET was hampered in performing detailed studies of the $\gamma$-ray sky above
10~GeV, due to back-splash of secondary particles
produced by high-energy $\gamma$-rays causing a self-veto in the monolithic
anti-coincidence detector used to reject charged particles, and due to a non-calibrated detector response. GLAST, soon to be launched, will not be strongly affected by these effects since the
anti-coincidence shield was designed in a segmented
fashion (Moiseev et al. 2007). The effective area of the
GLAST will be roughly an order of magnitude larger than that of
EGRET leading to an increased sensitivity for detecting celestial $\gamma$-ray photons (see Fig. 1 of Funk et al. 2008).
GLAST's default observing plan is a survey mode
where the sky is uniformly  observed. 
The increase of sensitivity and the survey mode open the gate to the possible discovery of new phenomenology. 
This paper analyzes astrophysical scenarios that can be detected  at the upper end of the energy range of  GLAST, as a result of cosmic-ray (CR) diffusion in the ISM.

\section{The role of diffusion }

The $\pi^0$-decay $\gamma$-ray flux 
from a source of proton-density $n_{p}$ is given by
\be
   F(E_{\gamma})=2\int^{\infty}_{E_{\pi}^{\rm min}}
({F_{\pi}(E_{\pi})}/{\sqrt{E_{\pi}^{2}-m_{\pi}^{2}}})
   \;dE_{\pi},
\ee
{  where $E_{\pi}^{\rm min}$ is
the minimum pion energy 
given by $E_{\pi}^{\rm min}(E_{\gamma})=E_{\gamma}+ {m_{\pi}^{2}}/{4E_{\gamma}},$ and }
\be
F_{\pi}(E_{\pi})=4\pi n_{p}\int^{E^{\rm max}_{p}}_{E^{\rm
min}_{p}} J_p(E) ({d\sigma_{\pi}(E_{\pi},\;E_{p})}/{dE_{\pi}})
\;dE_{p},
\ee
with $d\sigma_{\pi}(E_{\pi},\;E_{p})/dE_{\pi}$ being the
differential cross-section for the production of $\pi^0$
{  of energy $E_{\pi}$ by a proton of energy $E_{p}$ in a $pp$ collision. 
For an study of different parameterizations of this cross section see, e.g., 
(e.g., Domingo-Santamaria \& Torres 2005, Kelner et al. 2006). The limits of integration in the last expression are obtained by kinematic considerations (e.g. Torres 2004).} 
We have implicitly neglected any possible gradient of CR or gas number density in the target.
The CR spectrum is  given by
$
   J_p(E,\;r,\;t)= [{c} \beta / {4\pi}] f,
$
where $f(E,\;r,\;t)$ is the differential number density of protons at
an instant $t$ and distance $r$ from the source. 
{  Just for comparison, the spectrum of  cosmic-rays in the Earth vicinity is (e.g., Dermer 1986)
$
   J_p(E)=2.2 \, (E/{\rm GeV})^{-2.75} \,{\rm cm^{-2} s^{-1} GeV^{-1} sr ^{-1} }.
$

The particle's differential number density} $f$ satisfies the radial-temporal-energy dependent diffusion equation 
\be
   ({\partial f}/{\partial t})=({D(E)}/{r^2}) ({\partial}/{\partial
   r}) r^2 ({\partial f}/{\partial r}) + ({\partial}/{\partial
   E}) \, (Pf)+Q,
\ee
where $P=-dE/dt$ is the energy loss rate of the
particles, $Q=Q(E,\;r,\;t)$ is the source function, and $D(E)$
is the diffusion coefficient, for which we assume a dependence on the particle's energy. The energy loss rate are due to ionization and nuclear interactions, which timescale is $\tau_{pp}$, with the latter dominating over the former for energies larger than 1 GeV. 
{  The nuclear loss rate is $P_{\rm nuc} = E/\tau_{pp}$, with $\tau_{pp}=(n_p\, c \, \kappa \, \sigma_{pp} ) ^{-1}$ being the timescale for the corresponding nuclear loss, $\kappa \sim 0.45$ being the inelasticity of the interaction, and $\sigma_{pp}$ being the cross section. }
 
Aharonian \& Atoyan (1996) presented a solution for the diffusion equation with an arbitrary diffusion coefficient, and impulsive  injection spectrum $f_{\rm inj}(E)$, such that  $Q(E,r,t) = N_0 f_{\rm inj}(E) \delta{\bar r} \delta(t)$. For the particular case in which $D(E)\propto E^\delta$ and $f_{\rm inj}\propto
E^{-\alpha}$, 
it reads
\be
  f(E, r,t) \sim ({N_0 E^{-\alpha}}/{\pi^{3/2} R_{\rm dif}^3}) \exp \left[ { - {(\alpha-1)t}/{\tau_{pp} }- ({R}/{R_{\rm dif}})^2} \right],
  \label{sol}
\ee
where \be R_{\rm dif} = 2 ( D(E) t [\exp(t \delta / \tau_{pp})-1]/[t \delta / \tau_{pp}])^{1/2} \ee stands for the radius of the sphere up to which the particles of energy $E$ have time to propagate after their injection.
In case of continuous injection of accelerated particles,  $ Q(E, \;
t)=Q_0 E^{-\alpha} {\cal T}(t) , $ the previous solution needs to be convolved with the function ${\cal T}(t-t')$ in the time interval $0 \leq t' \leq t$. 
{  If the source is described by a Heavside function,  
${\cal T}(t)=\Theta (t)$, 
 and for times $t$ less than the energy loss time, 
$
f(E,\;r,\;t)=({Q_0 E^{-\alpha}}/{4\pi D(E) r}) (
{2}/{\sqrt{\pi}})\int^{\infty}_{r/R_{\rm diff}} e^{-x^2}
dx,$}
(Atoyan et al. 1995).
We will assume typical values, $\alpha=2.2$ and $\delta=0.5$.

\begin{figure*}
\centering
\includegraphics[width=.45\columnwidth,trim=0 5 0 10]{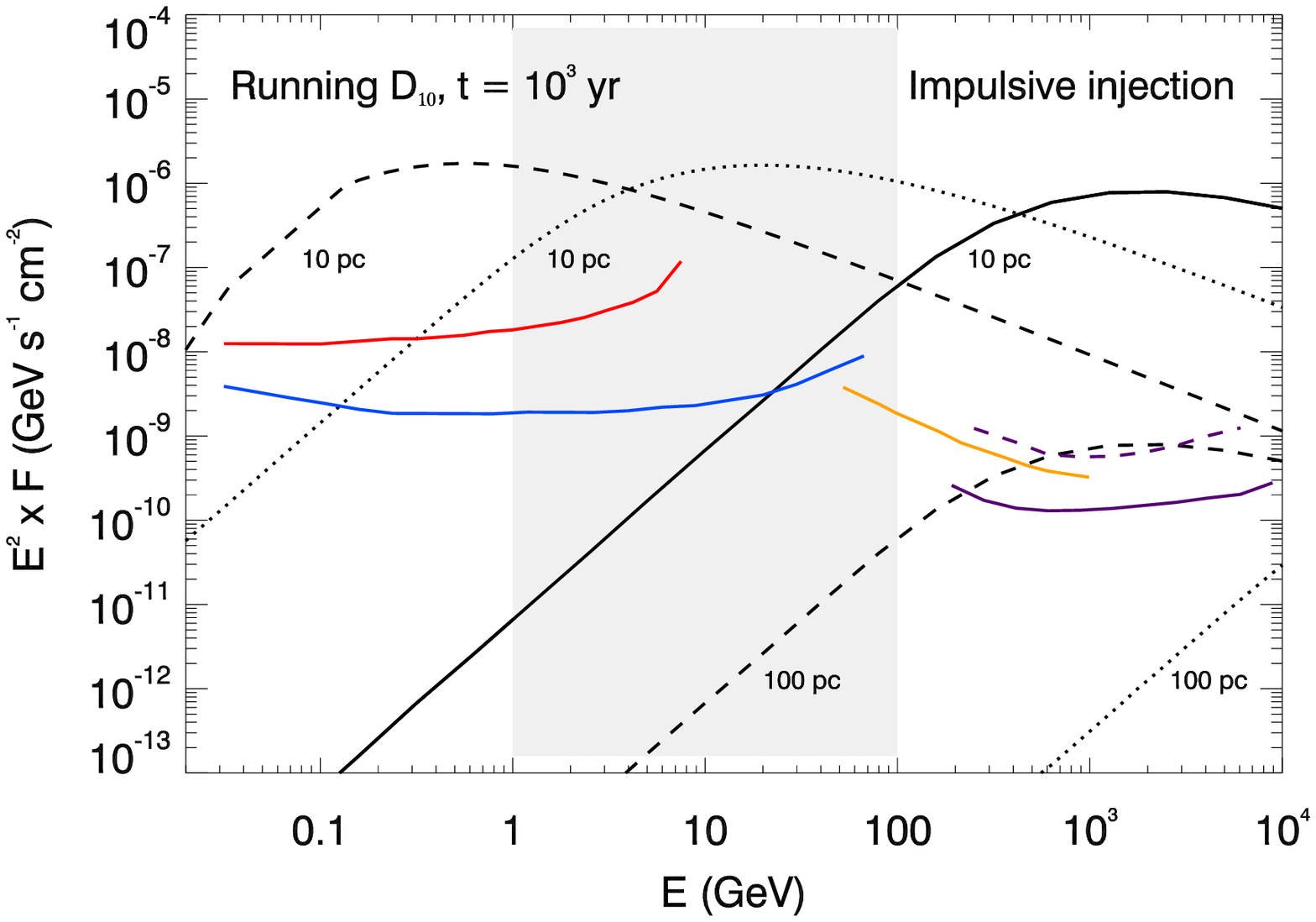}
\includegraphics[width=.45\columnwidth,trim=0 5 0 10]{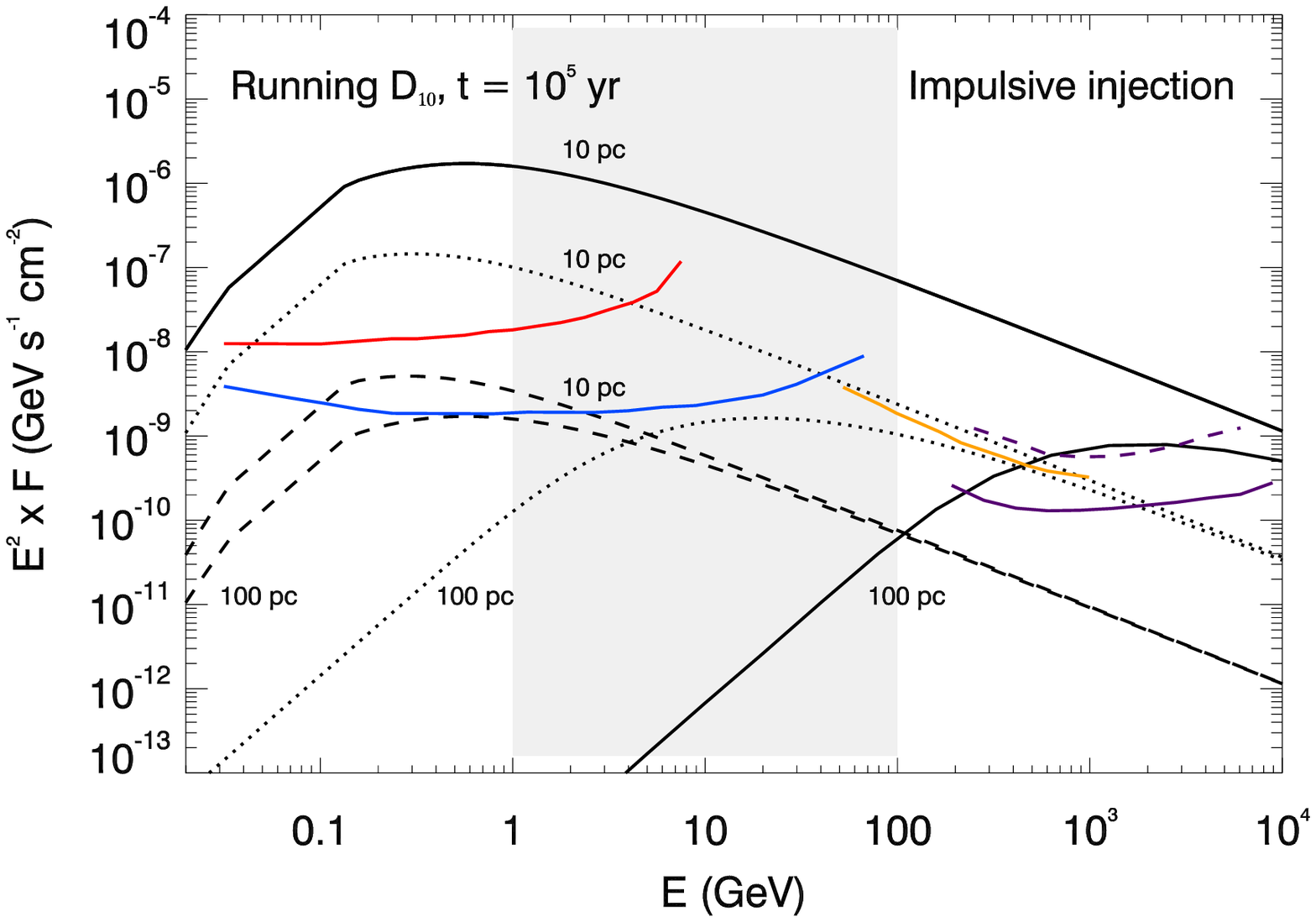}\\
\includegraphics[width=.45\columnwidth,trim=0 5 0 10]{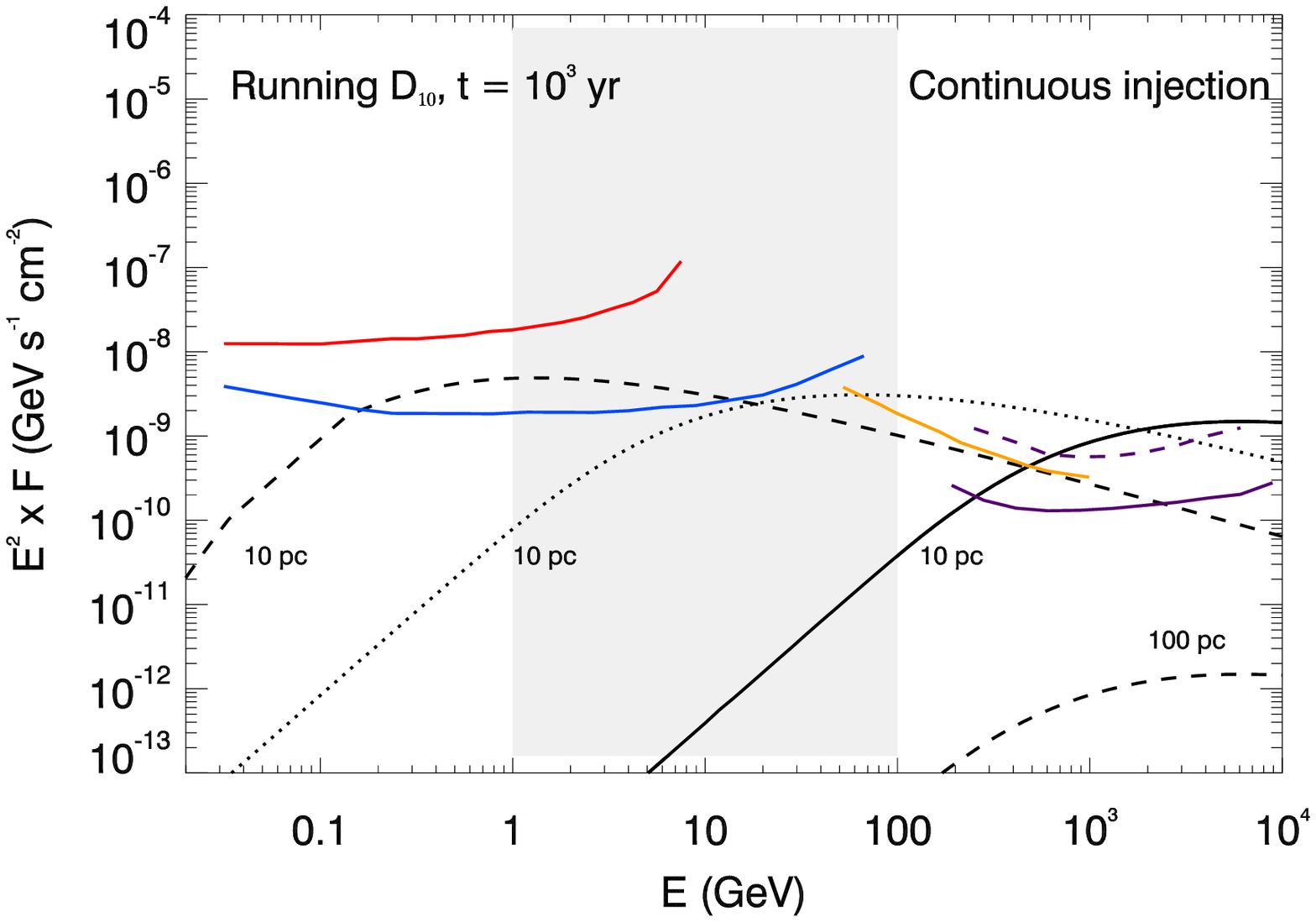}
\includegraphics[width=.45\columnwidth,trim=0 5 0 10]{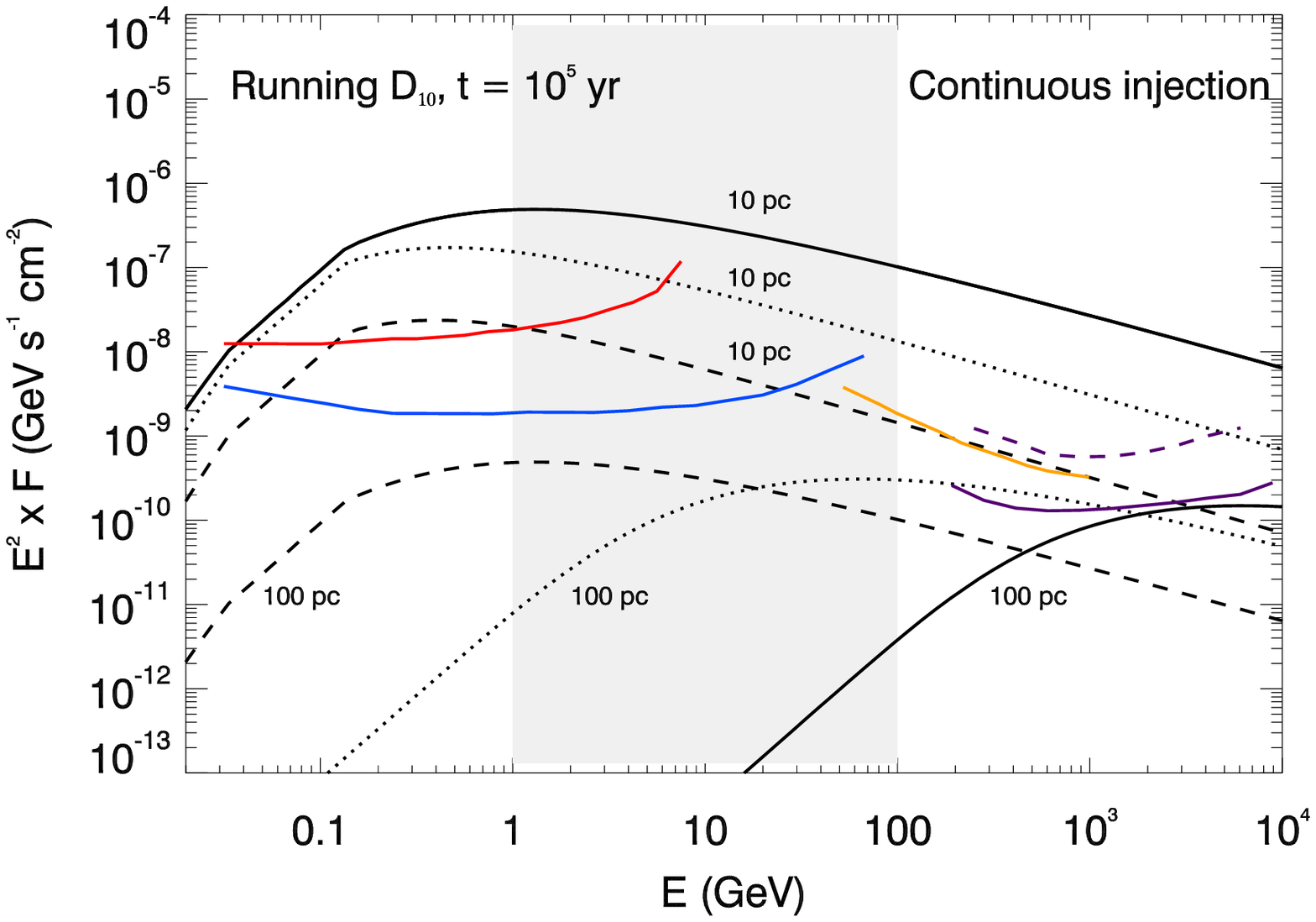}
\caption{SEDs generated by CR propagation in ISM with different properties.
Fluxes correspond to a cloud 
with $M_5/d^2_{kpc} = 0.5$. Curve for $D_{10}$ = 10$^{26}$,  10$^{27}$,  and 10$^{28}$ cm$^2$/s are shown with solid, dotted, and dashed lines respectively. Sensitivities of EGRET (red) and GLAST (blue), H.E.S.S. (magenta) (survey mode and pointed observations with typical integrations), and MAGIC (yellow), are shown for comparison (see Fig. 1 of Funk et al. 2008 for details on sensitivities).}
\label{tran-d10-ft}
\end{figure*}

In the case of energy-dependent propagation of CRs, a large variety
of $\gamma$-ray spectra is expected (e.g., Aharonian \& Atoyan 1996, Gabici \& Aharonian 2007, Torres et al. 2008). {  Diffusion of CRs have also been explored as an explanation for the high energy observations of the Galactic Center (e.g., Hinton \& Aharonian 2007).}
We have systematically studied, numerically producing over 2000 $E^2F$ distributions,  the dependences with the involved  parameters. 
Table \ref{depi} summarizes the results both for an impulsive and  a continuous accelerator. 
{  The influence of the total energy injected by the accelerator as CRs ($W_p$), the age of the accelerator ($t$), the medium density in which the CRs propagate ($n$), the diffusion coefficient of the medium (given at 10 GeV, $D_{10}$), the distance from the accelerator to the molecular cloud where the $\gamma$-rays are produced ($R$), the density of the cloud ($n_{Cl}$), its mass ($M_{Cl}$), and radius ($r_{cl}$), and the distance of such system to the observer ($d$) are described.}
The dominant dependences are related with the age of the accelerator and the diffusion coefficient.
{  These parameters both impact onto the CR distribution. The faster the diffusion, the farther the target can be from the accelerator, and still be subject to a significantly enhanced CR spectrum, }
see  Fig.  
\ref{tran-d10-ft}.
We use a source (cloud) parameterized in units of $M_5 = M_{Cl} / 10^5 M_\odot$ and $d_{kpc} = d / $ 1 kpc. 
$t_{transition}$, defined in the case of an impulsive accelerator, is the age for which the timescale for the corresponding nuclear loss becomes comparable to the age of the accelerator itself. 
$D_{transition}$ is the value of the diffusion coefficient for which the SEDs stop displacing in energy keeping approximately the same flux, as inferred from Fig.~\ref{tran-d10-ft}.

\begin{figure*}
\centering
\includegraphics[width=0.45\columnwidth,trim=0 5 0 10]{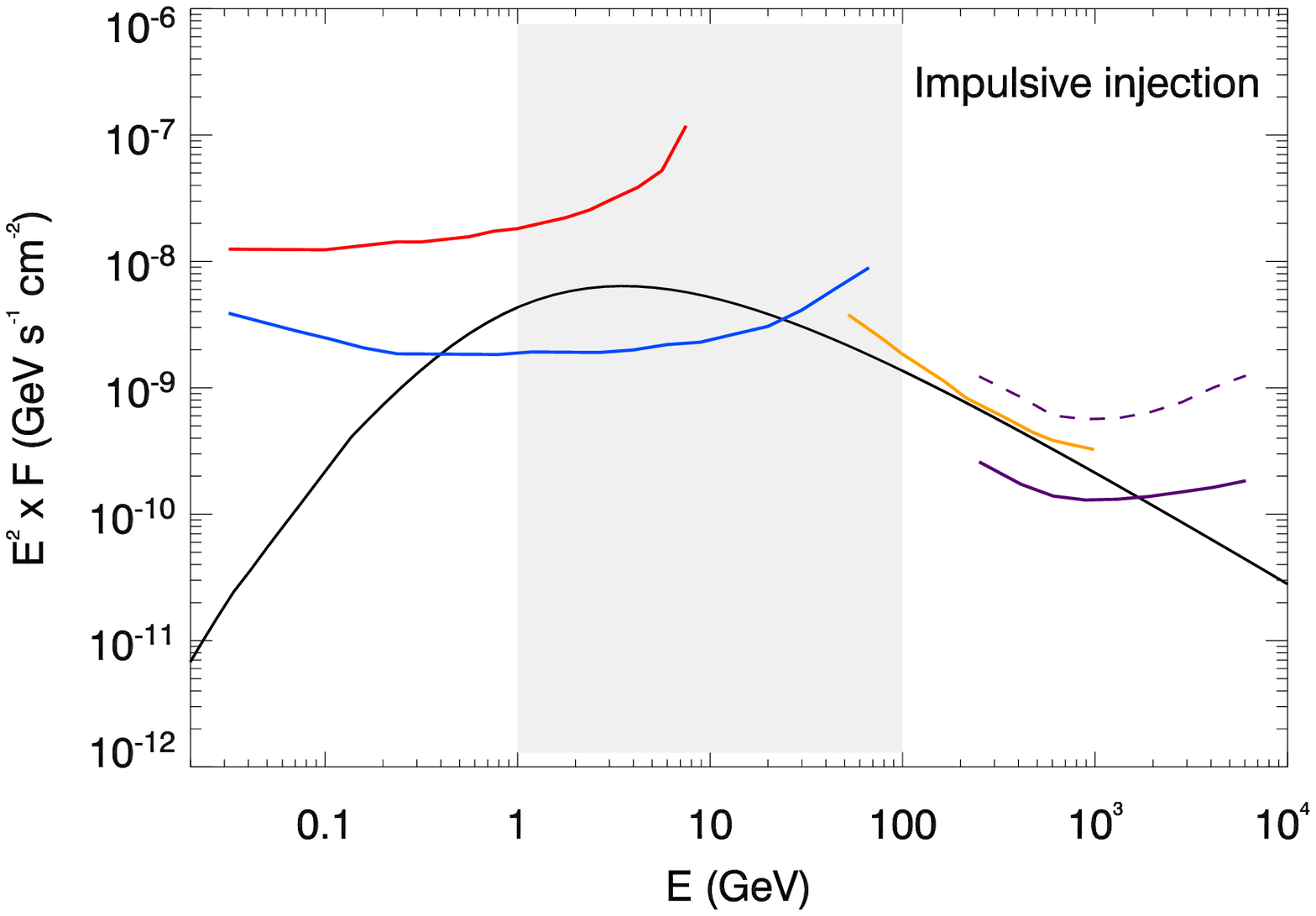}
\includegraphics[width=0.45\columnwidth,trim=0 5 0 10]{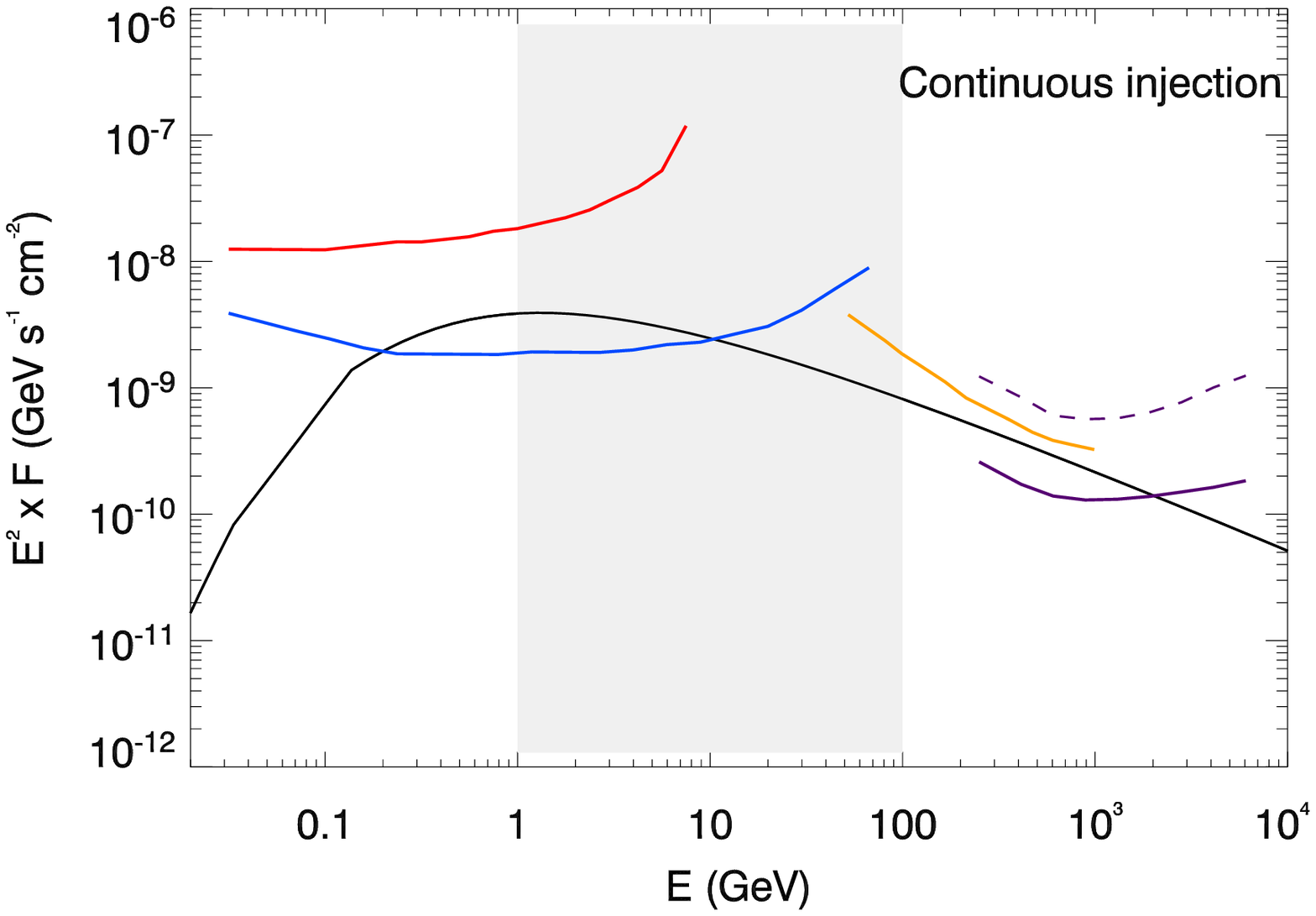}
\caption{
Examples of the model predictions for a hadronic maxima in the 1--100 GeV regime. 
The left panel shows the predictions for a cloud scaled at $M_5/d^2_{kpc} = 0.025$, located at 20 pc from an accelerator of $ 10^4$ yr, diffusing with $D_{10} = 10^{27}$ cm$^2$/s. The right panel curve shows the predictions for a cloud scaled at $M_5/d^2_{kpc} = 0.08$ located at 10 pc from an accelerator of $10 ^3$ yr, diffusing with $D_{10}$ = 10$^{28}$ cm$^2$/s. Increasing the ratio
$M_5/d^2_{kpc}$, the curves move up maintaining all other features. 
%
}
\label{had-zilla}
\end{figure*}

Setting, as an example,
reasonable parameters for the energy injected by the accelerator into
cosmic-rays (e.g., $W_p = 5\times10^{49}$ erg for an impulsive source and
$L_p = 5\times10^{37}$ erg/s for a continuous one) and for the interstellar
medium density (e.g., $n$ = 1 cm$^{-3}$), we have found several scenarios
for the appearance of hadronic maxima produced by diffusion. Examples are shown in
Fig.  \ref{had-zilla}, for the two types of accelerators. We have found that 
two kinds of peaks at this energy regime are possible:  those that are not to be detected by an instrument with the sensitivity of EGRET or MAGIC, and 
those that are not to be detected by an instrument like H.E.S.S. or VERITAS (the latter are not shown in the examples in Fig.  \ref{had-zilla}). We also note that the impulsive accelerator produces more steeper maxima. 
We find that maxima in the SED,  hadronically produced as an effect of diffusion of CRs, are possible and not uncommon at the high-energy end, where they produce $\gamma$-radiation at a level of flux detectable by the LAT.

\begin{figure*}[t!]
\centering
\includegraphics[width=.55\columnwidth,trim=0 5 0 10]{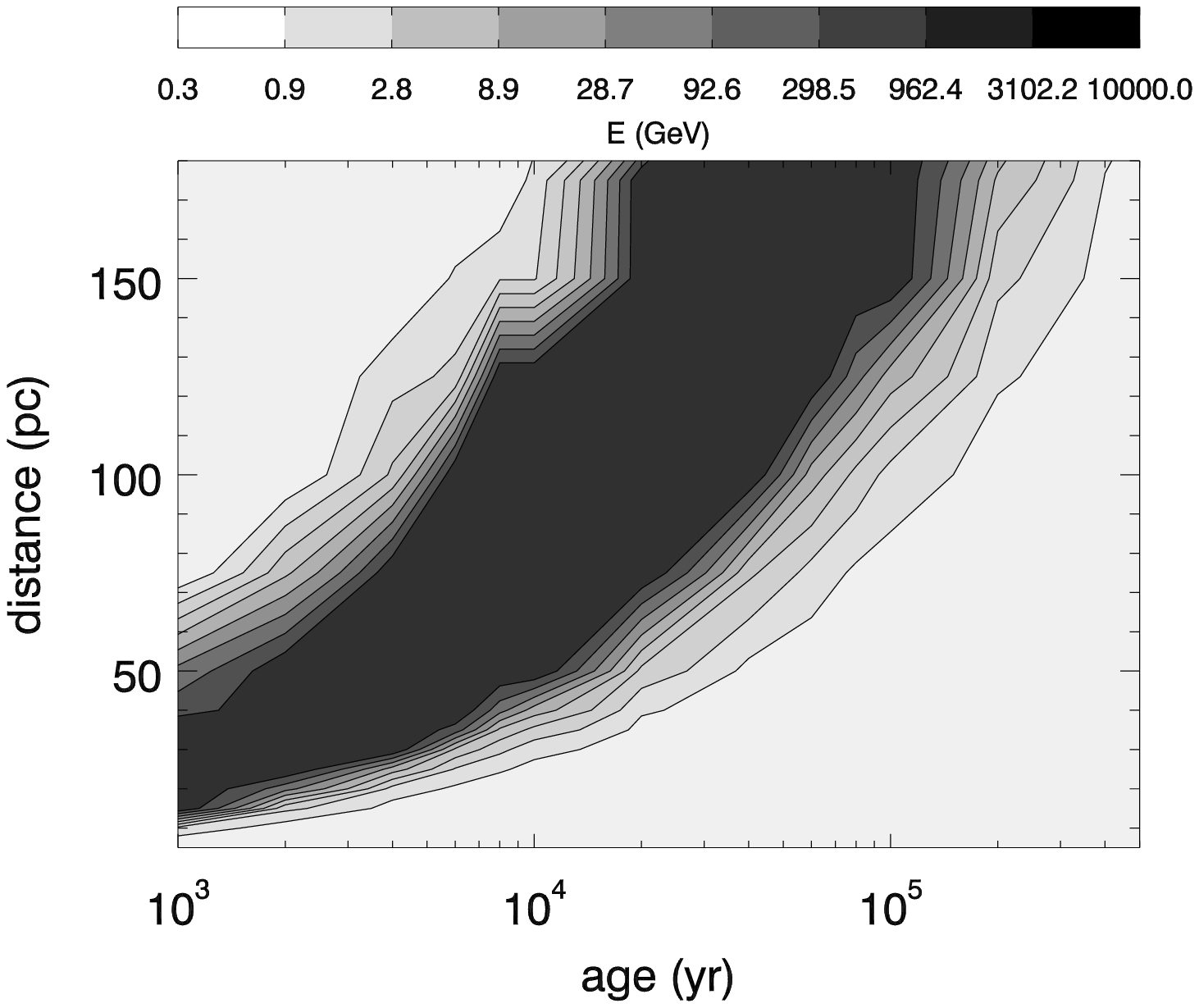}
\includegraphics[width=.55\columnwidth,trim=0 5 0 10]{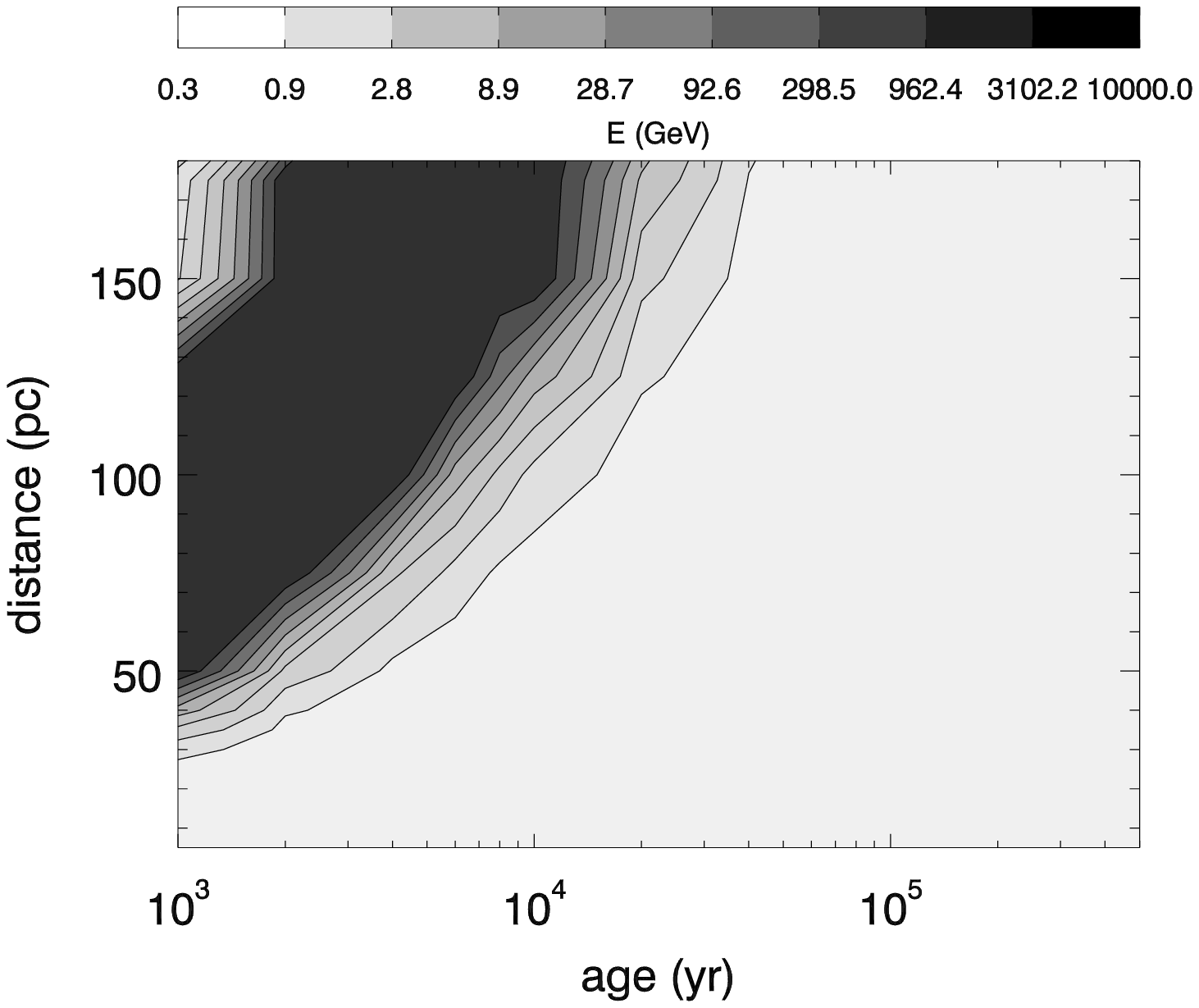}
\includegraphics[width=.55\columnwidth,trim=0 5 0 10]{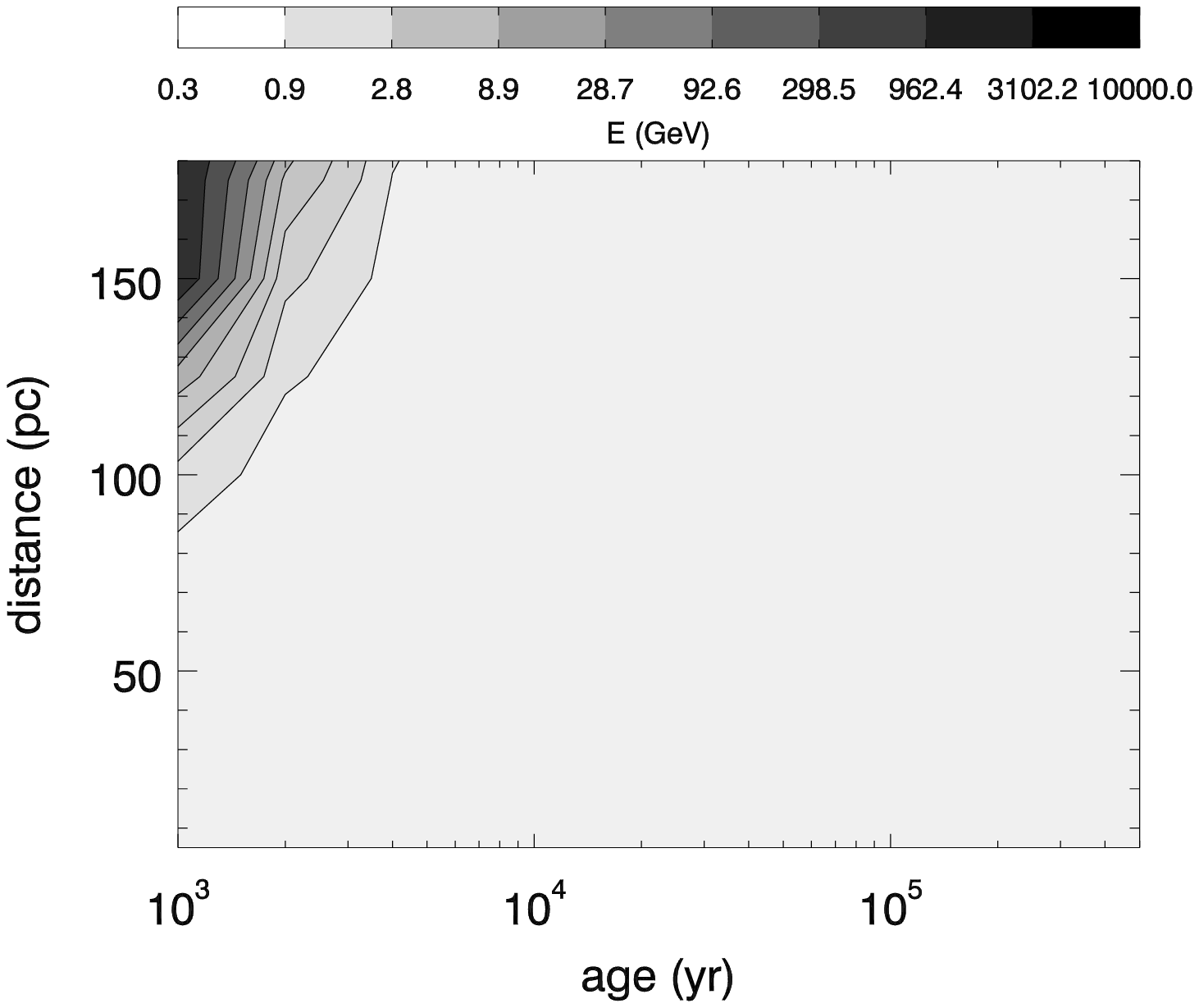}
\caption{
For each combination of age and accelerator-target separation, for which more than two thousand spectra where numerically produced,  the energy of the maximum of such spectra are shown in a contour plot. The color of the different contours corresponds to the range of energy where the maximum is found according to the color bar on the top of the figures. From left to right, plots are created for the case of an impulsive source injecting protons in a medium with $D_{10}$=$10^{26}$cm$^2$/s, $10^{27}$cm$^2$/s 
and $10^{28}$cm$^2$/s. }
\label{obs-1}
\end{figure*}

Fig. \ref{obs-1} shows, as contour plots, the energy at which the maximum of the SED is found for the cases of impulsive acceleration of cosmic rays, at different distances, ages of the accelerator, and diffusion coefficient. The energy-dependent propagation effects underlying our expectations for the diffusion of CRs in the studied parameter space
are clearly depicted there. The diffusion radius,  for $t \ll \tau_{pp}$, is $R_{\rm dif} (E) = 2 \sqrt{D(E) t}$, so that at a fixed age and distance, only particles of higher energy will be able to compensate a smaller $D_{10}$, producing SED maxima at higher $E$-values.   
The smaller values of $D_{10}$ that we study are expected in dense regions of ISM (e.g., Ormes et al. 1988, Torres et al. 2008).
It is interesting to note that for many, albeit not for all, of the SEDs studied, the maximum in $E^2F$ space is found at energies beyond the energetic range of GLAST. On the other hand, we also note that the use of Fig. \ref{obs-1} for the the interpretation of 
a GLAST observational discovery of a 1 -- 100 GeV maximum, provides interesting clues about the nature of the astrophysical system that generates the $\gamma$-rays. First, we find these SEDs in cases where the scenario does not predict
detectable emission at the EGRET sensitivity, so that they will represent new phenomenology. 
Second, we see from Fig.  \ref{obs-1} that the range of accelerator-target separations and ages of accelerator that would produce such a 1--100  GeV maximum is rather limited (see in Fig. \ref{obs-1} the narrow contours for maxima at such energies), which would lead to a direct identification of the source, in case  such system is found in the vicinity of the GLAST detection.

\begin{figure*}
\centering
\includegraphics[width=.45\columnwidth,trim=0 5 0 10]{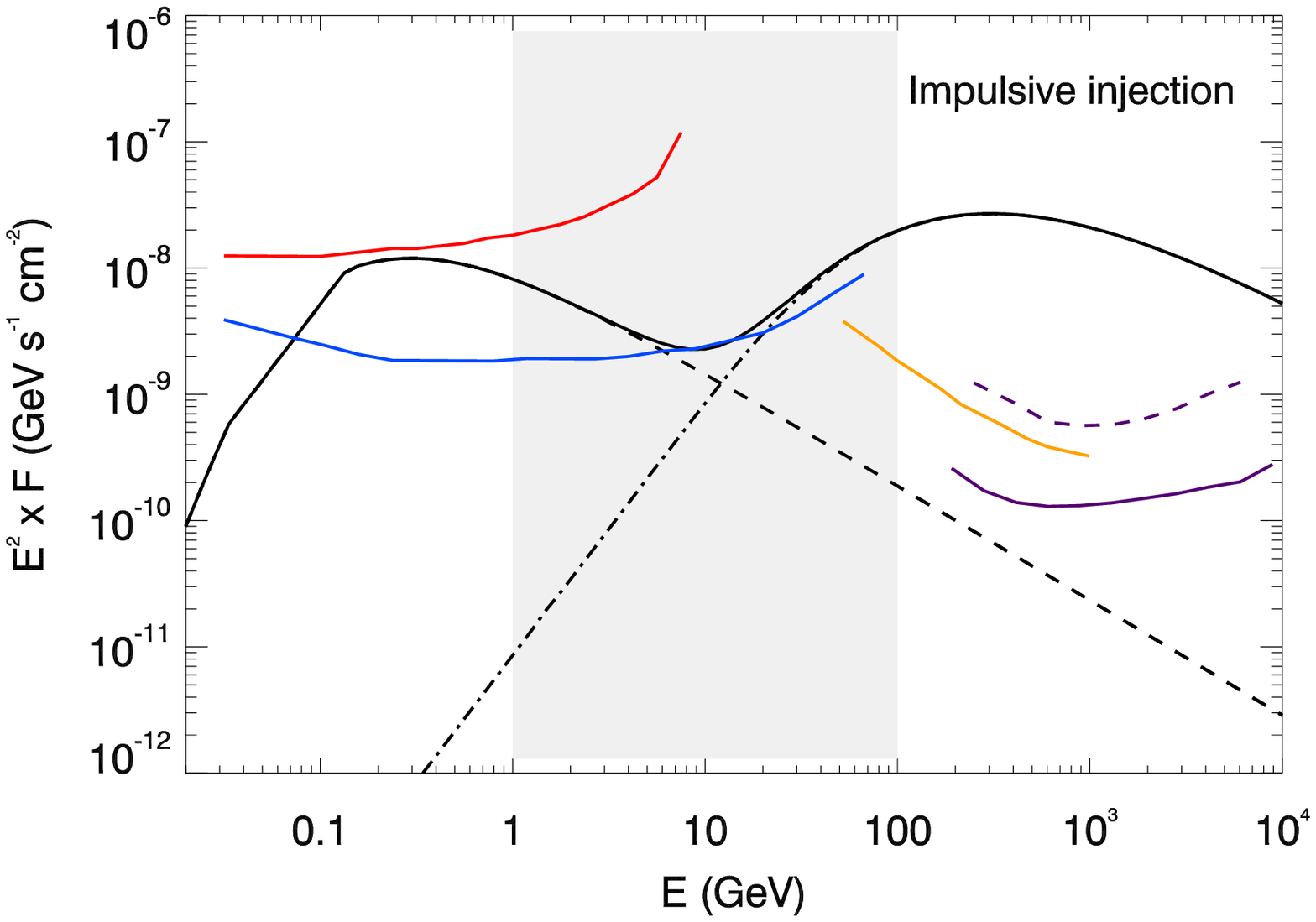}
\includegraphics[width=.45\columnwidth,trim=0 5 0 10]{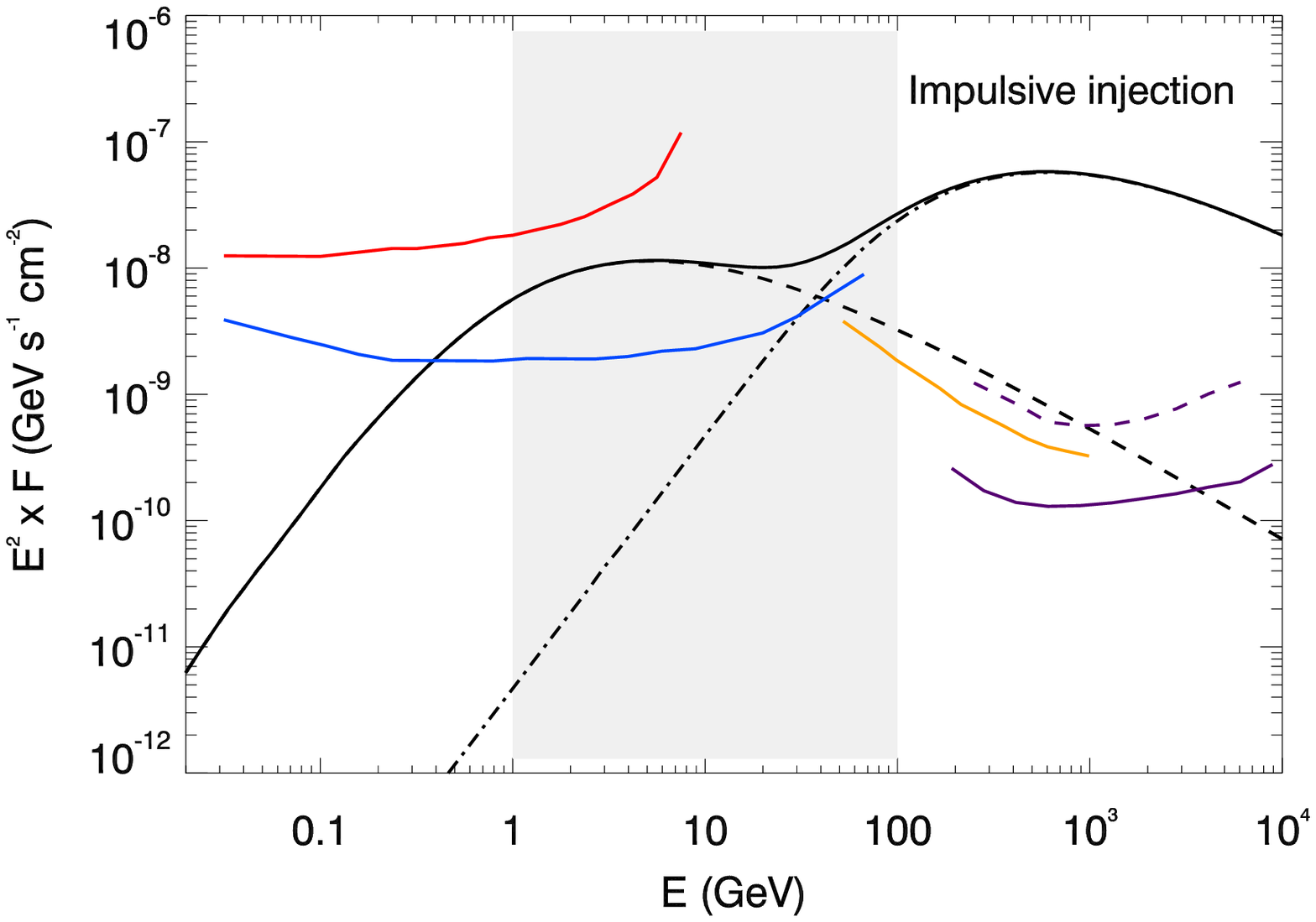}\\
\includegraphics[width=.45\columnwidth,trim=0 5 0 10]{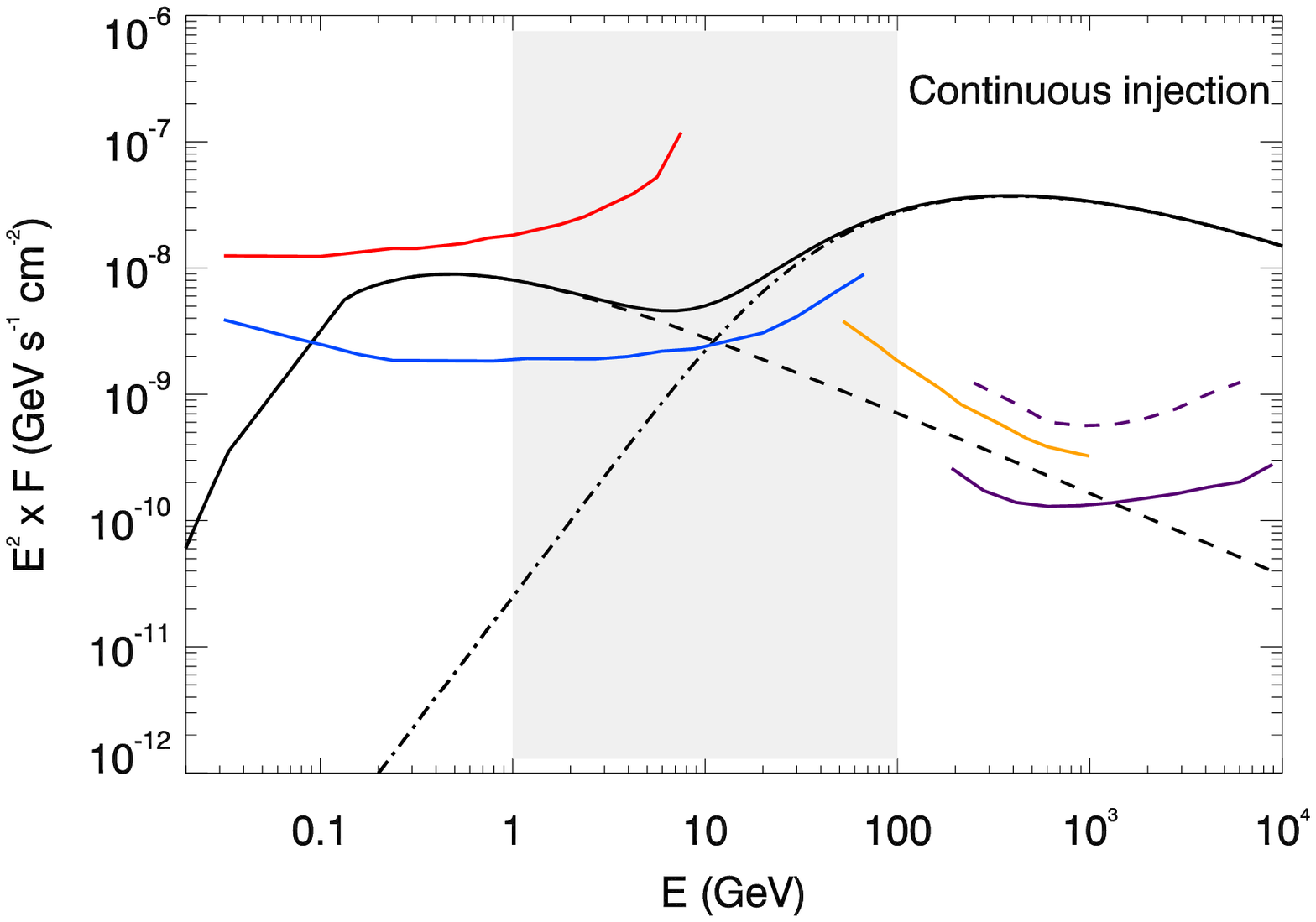} 
\includegraphics[width=.45\columnwidth,trim=0 5 0 10]{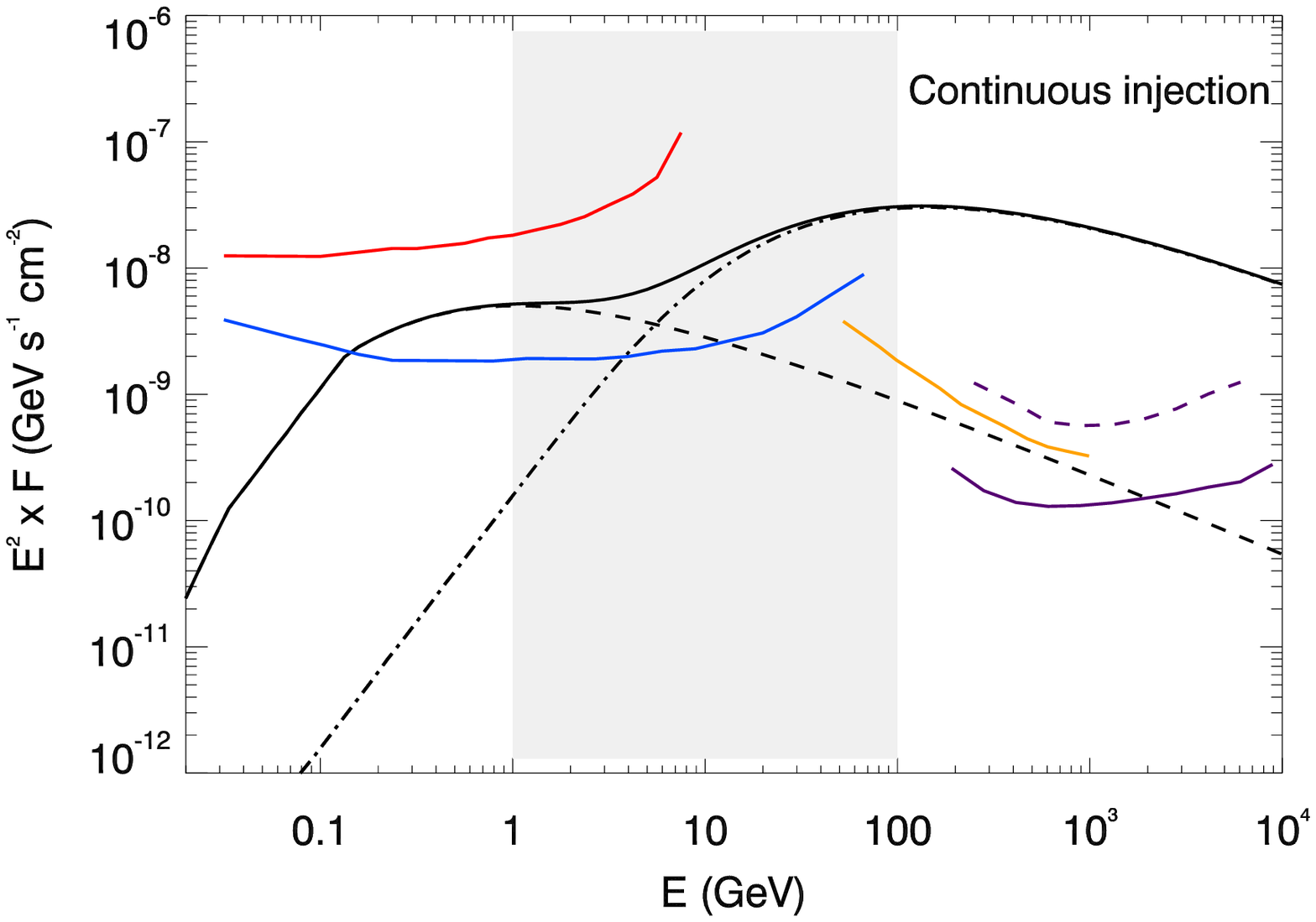}
\caption{The parameters for the plots are as follows, (top left) the dashed curve on the left: $t$ = 4$\times$10$^5$ yr, $R$
= 5 pc, $M_{5}/d^2_{kpc}$ = 0.01; the dashed curve on the right: $t$ =
10$^4$ yr, $R$ = 20 pc, $M_{5}/d^2_{kpc}$ = 0.1; (top right) the dashed curve
on the left: $t$ = 2$\times$10$^6$ yr, $R$=100 pc, $M_{5}/d^2_{kpc}$ = 3; the
dashed curve on the right: $t$ = 4$\times$10$^3$ yr, $R$ = 15
pc,$M_{5}/d^2_{kpc}$ = 0.1; (bottom left) the dashed curve on the left: $t$ = 2$\times$10$^6$ yr,
$R$=15 pc, $M_{5}/d^2_{kpc}$ = 0.004; the dashed curve on the right: $t$ =
10$^3$ yr, $R$ = 5 pc, $M_{5}/d^2_{kpc}$ = 1; ( bottom right) the dashed curve on
the left: $t$ = 2$\times$10$^6$ yr, $R$ = 40 pc, $M_{5}/d^2_{kpc}$ = 0.017; the
dashed curve on the right: $t$ = 6$\times$10$^4$ yr, $R$ = 30
pc,$M_{5}/d^2_{kpc}$ = 2.5. $D_{10}$ is set to
10$^{26}$ cm$^2$/s. $R$ is the accelerator-cloud separation.}
\label{V}
\end{figure*}

We now focus on the case of one unresolved but composite GLAST source. We thus consider two separate accelerator-cloud complexes that are close to the line of sight such that GLAST observe them as a single source, within its PSF. This kind of scenarios would yield the observational signature of an inverted spectrum.
Fig. ~\ref{V} shows four possible inverted spectra. The two figures in the top (bottom) part are generated by an impulsive (continuous) accelerator. The SED characterizing the oldest (youngest) accelerator is shown by dashed (dot-dashed) lines in each of the scenarios. Even if  EGRET could have been able to weakly detect some the inverted 
source models we simulated, it could not conclusively relate it to such phenomenon due to its large low-energy PSF. 
The counterpart at higher energies is a bright source potentially detectable by ground-based telescopes. Due to continuous energy coverage, GLAST is a prime instrument to track this phenomenology. 
The right panel cases show particular examples in which the detection of the source by an instrument with the sensitivity of EGRET is not possible at all. The inverted spectrum is less deep in these scenarios. Less pronounced V-shaped spectra can be obtained with concomitantly lower fluxes at TeV energies.

\section{Solution validity and timescales}

{  

Essentially, if we use the equations given above to compute $\gamma$-ray fluxes, we are assuming that 
there is no significant cosmic-ray gradient in the target (i.e., the gamma-ray emissivity is constant throughout the cloud). This assumption may be valid when the size of the cloud is less than the distance to the accelerator, and the diffusion coefficient inside and outside the cloud are not significantly different (or even if they are, the proton-proton timescale is larger than the time it takes for cosmic-ray to overtake the whole cloud).

Here we did not parameterize on the mass of the cloud, but rather on the value of $M_5$/$d_{kpc}^2$, i.e., the mass of the cloud in units of 10$^5$ M$_\odot$ divided by the distance squared in kpc, which we now call $A$ for brevity.
In order for the solutions to apply, then, two conditions need to be satisfied, $\tau_{pp} > t$ and the size of the cloud $r_{cl}$ (or simply $r$ below) has to be such $r_{cl}<R$ (see Table 1 to refresh the meaning of variables, and do not confuse $r$, the size of the cloud, with $R$, the separation between the cloud and the accelerator). 
The density of the cloud, presented in terms of $A$, is
$ 
n_{cl}=10^6  A d_{kpc}^2 / r_{pc}^3 \; {\rm cm}^{-3}.
$
The nuclear loss rate is 
$
\tau_{pp} \sim 100 r_{pc}^3 / (A  d_{kpc}^2) \; {\rm yr}.
$
The second condition (i.e, $r<R$) would be immediately satisfied if, say, $r = R/x$, with $x>1$. Then, an astrophysical scenario (not unique!) which fulfills also the first condition ($\tau_{pp} > t$) too is found when 
$
B=\sqrt{(100/A) \, (R^3/(x^3 t))} > d_{kpc}.
$ 
To be viable, of course $d_{kpc} < 8$ kpc, what can be used to define a minimum value of $x$, given an specific model with fixed $A$, $R$, and $t$. A final check can be done by comparing that the density obtained by replacing the previous inequalities, 
$
n_{cl}\, < \, (10^8/t_{yr})\;   {\rm cm}^{-3}
$
is in the range found in molecular clouds (which for the ages considered, is always fulfilled).
Table 2 shows the configurations used in the different figures of our work, 
together with the minimum value of $x$ such that $r<R$, $\tau_{pp} > t$, and $d_{kpc} < 8$ are all maintained and the solutions used applicable.

It is also interesting to discuss the different timescales within the cloud (see Gabici et al. 2007). In the former work, 
it was shown that  dynamical and advection cloud timescales do not play a relevant role in this problem. To estimate the degree of cloud penetration by cosmic ray it suffices to compare the loss and diffusion timescales. The former was derived above. The latter can be written as 
\be
\tau_{cl-dif} = 725 \xi \left( \frac{10^{27}}{D_{10}} \right)
\left( \frac{r_{pc}}{5} \right)
\left( \frac{E}{{\rm GeV}} \right)^{-0.5}
\left( \frac{B}{10\mu{\rm G}} \right)^{0.5}
\ee
where the diffusion coefficient inside the cloud has been parameterized as 
\be
D_{cl}(E) = 3.1 \xi D_{10} \left( \frac{E/GeV}{B / \mu{\rm G}} \right)^{0.5}.
\ee
In here, $\xi < 1 $ would account for a possible suppression of the diffusion coefficient inside the cloud as compared with that of the environment. 
For the proton-proton timescale to be larger than the time it takes for cosmic-ray to overtake the whole cloud, following the notation above, and taking an average cloud magnetic field of 10 $\mu$G, the following condition applies
\be
\frac{R}{x\,A\,d_{kpc}^2} \; \frac{D_{10}}{10^{27} {\rm cm^2 \, s}^{-1} } >  \frac{0.29}{\xi} 
\left( \frac{B / 10 \mu{\rm G}} {E/GeV} \right)^{0.5}.
\ee
With the parameters summarized in Table 2, it can be shown that plenty of clouds exist such that the 
diffusion timescale is shorter than the energy loss one at all energies, and especially at the ones we focus at the higher end of GLAST, and so we can neglect this effect. Our values of $A$ and typical galactic distances make in general for not so massive clouds. Our values of $R$ make in general for not so large clouds. In this situation, only for low diffusion coefficient (e.g., 10$^{26}$ cm$^2$ s$^{-1}$) timescales could become comparable if $\xi <1$; but then again, 
no significant suppression is expected when the environment has already such a low 
$D_{10}$.

}

\section{Concluding remarks}

Compton peaks (of which the first example could have been found already, Aharonian et al. 2006) are not the only way to generate a maximum in a SED in the range of 1--100 GeV.
A large variety of parameters representing physical conditions in the vicinity of a CR accelerator could produce a rather similar effect. Distinguishing between these cases would require multiwavelength information, search for counterparts, and modeling. If such a maximum is interpreted hadronically, as a result of diffusion of CR in the ISM and their subsequent interaction with a nearby target, the results herein presented constrain, given the energy at which the maximum of the SED is reached, the characteristics of the putative accelerator, helping to the identification process.
Indeed, one of the most distinguishing aspects of this study
is the realization that these signatures (in particular, a peak at the 1--100 GeV energy region) is indicative for an identification of the
underlying mechanism producing the $\gamma$-rays that is realized by nature: which accelerator (age and relative position to the target cloud) and under which diffusion properties CR propagate, as it is exemplified in Figure 3. 
In a survey mode such as the one GLAST will perform, it might also be possible to observe rather unexpected, tell-tailing  SEDs, like those V-shaped presented here, if observed with instruments having a limited PSF, predictably leaving many Galactic sources unresolved. 
Indeed, 
we finally remark about the V-shaped spectra presented here that we have focused in the situation where {\it from the same place in the sky} we have a double peak structure.
If  instead two nearby sources at different energies are indeed resolvable by GLAST, i.e., displaced already in the
GLAST range, and both above the threshold for detection,
we would have a clear case of morphology change and spatially dislocation with energy what would certainly make the study also interesting. 
As noted recently \footnote{In a poster by Gabici, Aharonian and Casanova in the Gamma-2008 meeting, July 6-11, 2008,  Heidelberg, and to our knowledge yet unpublished.}, a double peak could also be seen -coming from the same place in the sky- if we consider two populations of CRs interacting with the same cloud, for instance, the usual $E^{-2.75}$ Galactic bath of cosmic rays (producing a photon-spectrum peaking in the GeV regime) and the escaped CRs from the nearby source (producing a photon spectrum peaking in the TeV range). This possibility would, however, be applicable only to the case of very young accelerators, e.g., 2000 years or so, with clouds significantly separated from it, and thus it is a less general scenario than what we have explored in this paper as the mechanism for V-spectra production.

\acknowledgements
We acknowledge  support by grants MEC-AYA 2006-00530, CSIC-PIE 200750I029, and FPI BES-2007-15131, and A. Sierpowska-Bartosik, S. Funk, and J. Hinton for discussions.  A. Y. R. M. acknowledges Stanford University for hospitality during a visit.


\clearpage

{\rotate
\begin{table*} 
\scriptsize
\caption{Dependence of the SED ($E^2F$ vs. $E$) on the different  parameters. Imp. (cont.) stands for the impulsive (continuous) accelerator case. Dependences upon cloud parameters such as density ($n_{Cl}$), mass ($M_{Cl}$), and radius ($r_{cl}$) are obvious and related.}
\vspace{0.1cm}
  \centering
  \begin{tabular}{ ll}
    \hline
    parameter symbol and meaning & effect on the $E^2F$ distributions versus $E$ \\ \hline \hline
    {  Accelerator} &  \\
    \hline 
    $W_p$: total energy injected for the accelerator as CRs& imp.: overall scaling, small effects in the range if in the typical range 10$^{50}$-10$^{51}$ erg  \\
  $L_p$:  energy injected per unit time & cont.: overall scaling, small effects in the range if in the typical range 10$^{37}$-10$^{38}$ erg/s  \\
    increasing $t$: age of the accelerator & imp.: peak displaces to smaller energies for a fixed distance, \\
    & until $t>t_{transition}$, and the peak displaces to smaller fluxes \\
      & cont.:  peak displaces to smaller energies and larger fluxes, for a fixed distance \\ 
    \hline \hline
    {  Interstellar medium} &  \\
    \hline 
    $n$: density & negligible effects in the typical range 0.5-10 cm$^{-3}$, since $\tau_{pp} \gg t$. \\
    increasing $D_{10}$: & {  for a fixed age}: curve displaces to smaller energies until $D_{10} > D_{transition}$ \\
             the diffusion coefficient of the medium (at 10 GeV) & where peaks generated by clouds at large separation, $R$, displace up \\
             & and peaks generated by clouds at smaller $R$ displace down in the SED \\   
         & {  for a fixed distance}:  curve displaces to smaller energies until $D_{10} > D_{transition}$ \\
        & where peaks generated by older accelerators (larger $t$) displaces down \\ 
        & and peaks generated by younger accelerators (smaller $t$) displaces up \\    
    \hline \hline
  \end{tabular}
   \label{depi}
\end{table*}}

\begin{table} 
\label{mm}
\scriptsize
\caption{Configurations used in this work and  validity of the CR solutions used. }
\vspace{0.1cm}
  \centering
  \begin{tabular}{llll  |   l}
    \hline
Model         & $M_5$/$d_{kpc}^2$ & $R_{pc}$ & $t_{yr}$ & $x_{min}$ \\
         \hline
Fig.  1               & 0.5                          & 10    & 10$^3$     &  1.4 \\
Fig.  1               & 0.5                          & 10    & 10$^5$     &  0.3 \\
Fig.  1               & 0.5                         & 100    & 10$^3$    &  14.6\\
Fig.  1               & 0.5                            & 100    & 10$^5$  & 3.1 \\
Fig.  2 left         & 0.025                      & 20            & 10$^4$ & 3.6   \\
Fig.  2 right       & 0.08                        & 10            & 10$^3$ & 2.7  \\
Fig.  4 top left   & 0.01                        & 5            & $4\times 10^5$  & 0.3  \\
Fig.  4 top left   & 0.1                          & 20            & $2\times 10^4$  & 1.8 \\
Fig.  4 top right   & 3                           & 100            & $2\times 10^6$ &  0.6  \\
Fig.  4 top right   & 0.1                        & 15            & $4\times 10^3$  & 2.4  \\
Fig.  4 bottom left   & 0.004                 & 15            & $2\times 10^6$  &  0.9  \\
Fig.  4 bottom left    & 1                       & 5            & $4\times 10^3$  &  0.4   \\
Fig.  4 bottom right   & 0.017               & 40            & $2\times 10^6$  &  1.4  \\
Fig.  4 bottom right   & 2.5                   & 30            & $6\times 10^4$  & 0.7  \\
\hline
  \end{tabular}
\end{table}

\clearpage

\end{document}